\documentclass[useAMS,usenatbib,usegraphicx]{mn2e}

\title[Berkeley 58 and CG Cas]{Galactic clusters with associated Cepheid variables. VII. Berkeley 58 and CG Cassiopeiae}
\author[Turner et al.]
{D.~G. Turner$^{1,9,10,11}$\thanks{Email: turner@ap.smu.ca}, D. Forbes$^{2,9}$, D. English$^{2,10}$, P.~J.~T. Leonard$^{3}$, J.~N. Scrimger$^{4}$, 
\newauthor A.~W. Wehlau$^{5}$, R.~L. Phelps$^{6,9}$, L.~N. Berdnikov$^{7,11}$ and E.~N. Pastukhova$^{8}$ \\ \\
  $^1$Department of Astronomy and Physics, Saint Mary's University,
    Halifax, NS B3H 3C3, Canada \\
  $^2$Department of Physics, Sir Wilfred Grenfell College, Memorial University of Newfoundland, 
    Corner Brook, NF A2H 6P9, Canada \\
  $^3$ADNET Systems, Inc., 7515 Mission Dr., Suite A1C1, Lanham, Maryland 20706, U.S.A. \\
  $^4$Faculty of Computer Science, Dalhousie University, Halifax, NS B3H 1W5, Canada \\
  $^5$Department of Physics and Astronomy, The University of Western Ontario, London, ON N6A 3K7, 
    Canada \\
  $^6$Office of Integrative Activities, National Science Foundation, Division of Astronomical 
    Sciences, 4201 Wilson Blvd., Arlington, \\
    Virginia 22230, U.S.A. \\
  $^7$Sternberg Astronomical Institute, 13 Universitetskij prosp., Moscow 119992, Russia \\
  $^8$Institute of Astronomy, Russian Academy of Sciences, 48 Pyatnitskaya ul., Moscow 109017, Russia \\
  $^9$Visiting Astronomer, Kitt Peak National Observatory, National Optical Astronomy 
    Observatories, which is operated by the \\
    Association of Universities for Research in Astronomy, Inc. (AURA) under co-operative 
    agreement with the National Science \\
    Foundation \\
  $^{10}$Visiting Astronomer, Dominion Astrophysical Observatory, Herzberg Institute of Astrophysics,  
    National Research Council of Canada \\
  $^{11}$Visiting Astronomer, Harvard College Observatory Photographic Plate Stacks 
}
\date{Accepted 2008 May 1. Received 2008 May 1; in original form 2008 March 31}
\pagerange{\pageref{firstpage}--\pageref{lastpage}} \pubyear{2008}

\begin{document}

\label{firstpage}

\maketitle

\begin{abstract}
Photoelectric, photographic, and CCD $UBV$ photometry, spectroscopic observations, and star counts are presented for the open cluster Berkeley 58 to examine a possible association with the 4$^{\rm d}$.37 Cepheid CG Cas. The cluster is difficult to separate from the early-type stars belonging to the Perseus spiral arm, in which it is located, but has reasonably well-defined parameters: an evolutionary age of $\sim 10^8$ years, a mean reddening of E$_{B-V}{\rm (B0)}=0.70 \pm 0.03$ s.e., and a distance of $3.03 \pm 0.17$ kpc ($V_0-M_V=12.40 \pm 0.12$ s.d.). CG Cas is a likely cluster coronal member on the basis of radial velocity, and its period increase of $+0.170 \pm0.014$ s yr$^{-1}$ and large light amplitude describe a Cepheid in the third crossing of the instability strip lying slightly blueward of strip centre. Its inferred reddening and luminosity are E$_{B-V}=0.64 \pm 0.02$ s.e. and $\langle M_V \rangle =-3.06 \pm0.12$. A possible K supergiant may also be a cluster member.
\end{abstract}


\section{Introduction}
After the rediscovery in the early 1950s of spatial coincidences between Cepheids and open clusters by \citet{ir55,ir58}, Eggen \citep[see][]{sa58}, and \citet{kh56}, a number of searches for additional coincidences were made by \citet{kr57}, \citet{vb57}, and \citet{ti59}, among others. Tifft's search resulted in the discovery of a near spatial coincidence between the 4$^{\rm d}$.37 Cepheid CG Cassiopeiae and an anonymous open cluster, subsequently catalogued as Berkeley 58 \citep{sw62}, which lies less than one cluster diameter to the west. The field is coincident with a portion of the Perseus spiral arm that is relatively rich in open clusters, and the cluster NGC 7790 with its three Cepheid members lies in close proximity. The possibility that CG Cas might be an outlying member of 
NGC 7790 was raised at one time by \citet{ef64a,ef64b}, and found some support in a star count analysis by \citet{ko68}. More detailed star counts in the field \citep{tu85} indicate otherwise, as do the available proper motion data \citep{fr74,fr77}. The Cepheid does lie in the corona of Berkeley 58 \citep{tu85}, although Frolov has argued that it is not a probable cluster member.

Given a probable distance of ~3 kpc to both CG Cas and Berkeley 58 \citep[e.g.,][]{fr79,pj94}, it is not clear that existing proper motion data are precise enough to provide conclusive evidence pertaining to the cluster membership of CG Cas. The present study was therefore initiated in order to examine the case in more detail. As demonstrated here, there is strong evidence that CG Cas is a likely member of Berkeley 58 and that it can serve as a calibrator for the Cepheid period-luminosity (PL) relation.

\section{Observational Data}
A variety of observations were obtained for the present investigation. Table 1 presents photoelectric {\it UBV} photometry for bright members of Berkeley 58, obtained during observing runs at Kitt Peak National Observatory in 1981 September, 1982 August, and 1984 August. The data, acquired using 1P21 photomultipliers and standard {\it UBV} filter sets used in conjunction with pulse-counting photometers on the No. 4--0.4-m, No. 2--0.9-m, and 1.3-m telescopes at Kitt Peak, have associated uncertainties typical of our previous investigations of Cepheid clusters \citep{te92,tu92,te94}, namely standard internal errors for a single observation of $\pm 0.01$ in $V$ and {\it B--V}, and $\pm 0.02$ in {\it U--B}, for stars brighter than $V=13$. The estimated external errors for all but the faintest stars are similar in magnitude. The stars are identified by their numbering in Fig. 1, as well as by their 2000 co-ordinates in the 2MASS survey \citep{cu03}; the number of individual observations for each star is given in column 7 of Table 1.

\begin{figure}
\begin{center}
\includegraphics[width=6cm]{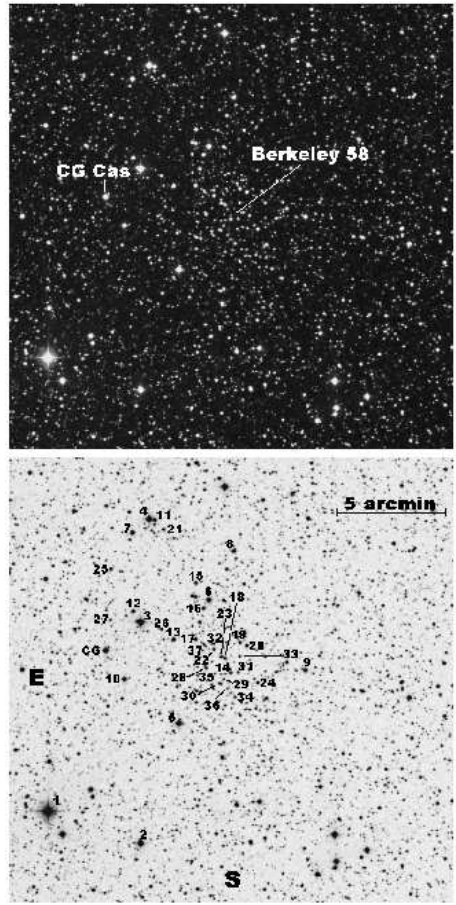}
\end{center}
\caption{A finder chart for the field of Berkeley 58 from the red image of the Palomar Observatory Sky Survey. The field of view measures $20^\prime \times 20^\prime$ and is centred at 2000 co-ordinates: RA = $00^{\rm h} 00^{\rm m} 12^{\rm s}.9$, DEC = +60$^{\circ}$ 56${\arcmin}$ 07${\arcsec}$. The top image depicts the location of CG Cas relative to the cluster core, the lower image identifies photoelectrically observed stars. [The National Geographic Society-Palomar Observatory Sky Atlas (POSS-I) was made by the California Institute of Technology with grants from the National Geographic Society.]}
\label{fig1}
\end{figure}

\setcounter{table}{0}
\begin{table*}
\begin{minipage}{12cm}
\caption[]{Photoelectric {\it UBV} Data for Stars in Berkeley 58.}
\label{tab1}
\begin{tabular}{cccccccl}
\hline
Star &RA(2000) &DEC(2000) &$V$ &$B-V$ &$U-B$ &n &Notes \\
\hline
CG Cas &00 00 59.24 &60 57 32.5 &11.20 &1.30 &+1.00 &15 &F5--G1 I \\
1 &00 01 21.61 &60 50 21.2 &7.28 &0.14 &$-0.39$ &4 & \\
2 &00 00 47.68 &60 48 49.8 &9.80 &0.27 &$-0.10$ &4 & \\
3 &00 00 46.10 &60 58 46.5 &9.85 &1.35 &+1.20 &4 & \\
4 &00 00 42.77 &61 03 26.1 &10.02 &0.58 &+0.02 &4 & \\
5 &00 00 32.75 &60 54 11.8 &10.79 &2.04 &+2.44 &2 &(K II)? \\
6 &00 00 20.63 &60 59 43.1 &10.95 &0.50 &$-0.12$ &4 &B3--5 V$^{\rm a}$ \\
7 &00 00 48.46 &61 02 49.5 &11.49 &0.46 &$-0.44$ &4 &B3--5 Vnn \\
8 &00 00 10.99 &61 01 53.6 &12.04 &0.59 &+0.26 &1 & \\
9 &23 59 45.42 &60 56 28.1 &12.11 &0.57 &+0.48 &3 & \\
10 &00 00 52.47 &60 56 14.1 &12.16 &2.25 &+2.51 &3 & \\
11 &00 00 40.07 &61 03 21.9 &12.35 &0.50 &$-0.25$ &4 &B2.5 V \\
12 &00 00 48.77 &60 59 17.2 &12.55 &1.41 &+1.10 &1 & \\
13 &00 00 33.91 &60 57 58.7 &12.78 &0.62 &+0.09 &4 & \\
14 &00 00 13.07 &60 56 25.4 &12.82 &0.57 &+0.04 &4 & \\
15 &00 00 25.25 &61 00 29.8 &13.12 &1.59 &+1.50 &3 & \\
16 &00 00 22.63 &60 59 20.7 &13.22 &0.52 &+0.21 &5 & \\
17 &00 00 25.83 &60 57 58.2 &13.30 &0.83 &+0.55 &4 & \\
18 &00 00 15.03 &60 57 05.0 &13.35 &0.57 &+0.03 &4 &B6 Vn \\
19 &00 00 09.46 &60 57 47.6 &13.36 &0.72 &+0.44 &2 & \\
20 &00 00 06.89 &60 57 37.6 &13.41 &0.60 &+0.11 &2 & \\
21 &00 00 36.95 &61 02 55.7 &13.41 &0.66 &+0.49 &3 & \\
22 &00 00 19.09 &60 57 29.8 &13.54 &1.55 &+1.39 &4 & \\
23 &00 00 16.44 &60 57 08.8 &13.60 &0.56 &+0.00 &4 &B5:: Vnn \\
23 &00 00 03.17 &60 55 57.1 &13.69 &0.57 &+0.10 &4 &B7 V \\
25 &00 00 56.70 &61 01 12.0 &13.71 &1.43 &+1.23 &2 \\
26 &00 00 38.39 &60 58 26.2 &14.11 &0.73 &+0.59 &4 & \\
27 &00 00 57.31 &60 58 54.9 &14.14 &0.84 &+0.26 &4 & \\
28 &00 00 22.18 &60 56 39.7 &14.20 &0.62 &+0.19 &5 & \\
29 &00 00 15.73 &60 56 08.6 &14.70 &0.57 &+0.18 &4 & \\
30 &00 00 16.73 &60 55 55.6 &14.71 &0.72 &+0.35 &5 &double \\
.. &... &... &15.46 &0.76 &... &CCD & \\
31 &00 00 10.33 &60 56 25.4 &14.74 &1.57 &... &2 & \\
32 &00 00 19.10 &60 57 44.5 &14.75 &0.62 &+0.44 &3 & \\
33 &00 00 09.64 &60 57 10.1 &14.91 &1.02 &+0.54 &1 & \\
34 &00 00 11.54 &60 55 19.5 &15.06 &1.09 &... &2 & \\
.. &... &... &14.88 &0.66 &+0.13 &CCD & \\
35 &00 00 18.88 &60 56 24.1 &15.09 &0.61 &+0.25 &4 & \\
37 &00 00 14.44 &60 55 43.3 &15.16 &1.17 &... &1 & \\
37 &00 00 23.28 &60 57 27.8 &15.63 &0.81 &+0.79 &3 & \\
\hline
\end{tabular}
$^{\rm a}$V654 Cas \citep{be93}.
\end{minipage}
\end{table*}

Star 6 is the eclipsing system V654 Cas, for which \citet{be93} cites photoelectric values of {\it V} and {\it B--V} outside of eclipse that are close to the values given here. Star 30 is a close optical double with components of nearly identical brightness. The photoelectric values apply to the combined light from both stars, whereas CCD observations provide uncontaminated data for the southwestern star of the pair, as established by its CCD magnitude being 0.75 mag. fainter. By contrast, the CCD {\it V} magnitude for star 35 is 0.21 mag. brighter, which suggests possible variability in the object. Individual photoelectric observations for CG Cas are presented in Table 2.

Photographic {\it UBV} photometry was also obtained for stars in the nuclear and coronal regions of Berkeley 58 from photographic plates of the cluster field obtained in 1984 September using the 1.2-m Elginfield telescope of the University of Western Ontario. The star images were measured using the iris diaphragm photometer at Saint Mary's University, and were reduced to the {\it UBV} system and calibrated with reference to the photoelectric standards identified in Table 1 using the techniques discussed by \citet{tw89}. The resulting data are presented in Table 3 (Appendix) in similar format to the data of Table 1, and the stars are identified by their 2000 co-ordinates. The photographic values for cluster stars in common with the CCD survey \citep{pj94} agree very closely with the CCD values, when the latter are adjusted to the present system. However, earlier photographic {\it UBV} photometry of cluster stars by \citet{fr79} displays systematic differences relative to the present data. Since the present survey samples a much larger number of cluster stars, no attempt was made to combine Frolov's data with the present photometry.

CCD {\it UBV} photometry for stars in the nuclear region of Berkeley 58 was published previously by \citet{pj94}, but for this study was recalibrated using the Table 1 stars as standards. The revised photometry for cluster stars is presented in Table 4 (Appendix), where the star numbers correspond to the scheme adopted by \citet{pj94}, incremented by 1000. The stars are also identified by their 2000 co-ordinates. Since the $U$ band measurements have a much brighter limit than the $B$ and $V$ measures, the CCD photometry is less useful for studying the reddening in the field. But it is valuable for identifying the faint portion of the cluster main sequence.

Spectroscopic imaging of bright stars in Berkeley 58 was made in 1984 July and 1985 September using the Cassegrain spectrograph on the 1.8-m Plaskett telescope of the Dominion Astrophysical Observatory. The observations, at a dispersion of 15 \AA\ mm$^{-1}$ and centred in the blue spectral region, were recorded photographically and later scanned for radial velocity measurement with the PDS microdensitometer at the David Dunlap Observatory of the University of Toronto \citep[see][]{td84}. It was also possible to estimate spectral types for the stars from the photographic spectra, with results presented in Table 1.

The field of the Cepheid CG Cas was also examined on archival images in the collections of Harvard College Observatory and Sternberg Astronomical Institute in order to obtain brightness estimates for the star and to construct seasonal light curves for comparison with a standard light curve constructed from photoelectric observations \citep{be07}. The resulting data were used to estimate times of light maximum for the Cepheid and to track its O--C changes, the differences between observed (O) and computed (C) times of light maximum. Rate of period change, in conjunction with light amplitude, is an excellent diagnostic of the location of individual Cepheids in the instability strip \citep{te06a}, such information providing an excellent parameter for comparison with what can be gleaned from information on the age of the surrounding stars provided by the cluster H-R diagram.

\section{Star Counts}
The first step in studying Berkeley 58 involved star counts made using a photographic enlargement from a glass copy of the POSS-E plate for the field. Strip counts in several different orientations delineated the cluster centre, followed by ring counts illustrated in Fig. 2; the centre of symmetry is located at RA = $00^{\rm h} 00^{\rm m} 12^{\rm s}.9$, DEC = $+60^{\circ} 56{\arcmin} 07{\arcsec}$ (2000). The upper portion of Fig. 2 illustrates ring counts for stars detected on the 2MASS survey \citep{cu03} to the survey limit , whereas the lower portion shows star counts from the Palomar Observatory Sky Survey (POSS) E-plate to two different magnitude limits.

\setcounter{table}{1}
\begin{table}
\caption[]{Photoelectric {\it UBV} Observations for CG Cassiopeiae.}
\label{tab2}
\begin{center}
\begin{tabular}{cccc}
\hline
HJD &$V$ &$B-V$ &$U-B$ \\
\hline
2444849.8938 &11.37 &1.28 &... \\
2444854.8655 &11.53 &1.36 &... \\
2444856.8539 &11.22 &1.14 &... \\
2444857.8358 &11.08 &1.14 &... \\
2444857.8689 &11.09 &1.16 &... \\
2445197.9177 &10.92 &1.04 &0.74 \\
2445205.9366 &11.45 &1.25 &0.84 \\
2445206.8851 &11.04 &1.10 &0.76 \\
2445933.8457 &11.73 &1.40 &1.04 \\
2445935.8748 &10.99 &1.08 &0.85 \\
2445937.8601 &11.59 &1.37 &0.96 \\
2445938.8420 &11.74 &1.38 &1.02 \\
2445939.8315 &10.85 &0.99 &0.72 \\
2445941.8773 &11.55 &1.35 &0.93 \\
2445942.7724 &11.76 &1.42 &1.04 \\
\hline
\end{tabular}
\end{center}
\end{table}

\begin{figure}
\begin{center}
\includegraphics[width=6cm]{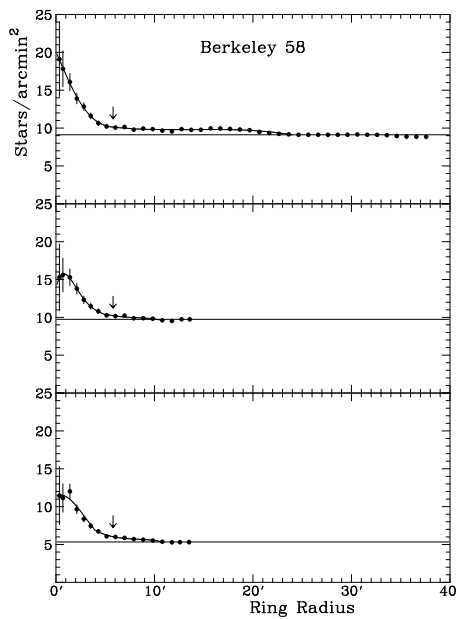}
\end{center}
\caption{Star densities for the field of Berkeley 58, as measured in rings relative to the adopted cluster centre. The upper diagram contains ring counts made from the 2MASS survey, the lower two diagrams ring counts from the POSS E-plate of the field for a faint limit (middle) and a brighter limit (lower). The location of CG Cas relative to the cluster centre is indicated by an arrow.}
\label{fig2}
\end{figure}

The counts from the 2MASS survey were made without regard for overlap with the star cluster NGC 7790, which lies $23\arcmin$ to the northwest of Berkeley 58, whereas the counts from the POSS-E plate were restricted beyond 11$\arcmin$ from the cluster centre to sectors that avoided overlap with the outlying regions of NGC 7790. The effect of contamination from the coronal region of NGC 7790 is detectable in the 2MASS star counts beyond roughly 12$\arcmin$ from the cluster centre, but because of restrictions imposed by the location of Berkeley 58 on the POSS, we were unable to establish uncontaminated star counts from the POSS-E plate beyond about 15$\arcmin$ from the cluster centre. Nevertheless, the two sets of counts appear to yield similar parameters for the inner regions of the cluster. Berkeley 58 is estimated to have a nuclear radius of $r_n \simeq 4\arcmin.5$ (4.0 pc) in the notation of \citet{kh69}, whereas the coronal (or tidal) radius is estimated to be $R_c \simeq 11\arcmin$ (9.7 pc) from the trends in the 2MASS star densities as well as the apparent flattening of the POSS-E star densities in the outermost rings.

Star counts predict a total of $197 \pm27$ members brighter than the limit of the 2MASS survey lying within 5$\arcmin$ of the cluster centre, $487 \pm82$ members within 11$\arcmin$ of the cluster centre, field stars within the same regions being 715 and 4835, respectively. Field stars clearly outnumber cluster members in both regions. CG Cas is located 5$\arcmin$.8 from the centre of Berkeley 58, in the cluster coronal region just beyond its nuclear boundaries. Although not projected on the core of Berkeley 58, CG Cas is spatially coincident with the cluster, which occupies most of the field of Fig. 1.

\begin{figure}
\begin{center}
\includegraphics[width=6cm]{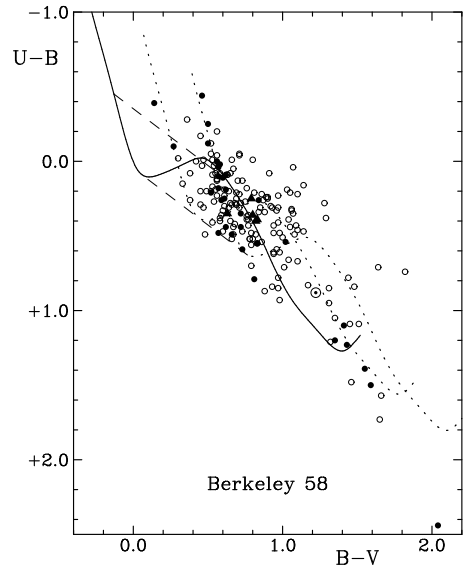}
\end{center}
\caption{A {\it UBV} colour-colour diagram for observed Berkeley 58 stars: photoelectric observations (filled circles), photographic observations supplemented by CCD observations (open circles), CCD observations (filled triangles), and CG Cas (circled point). The intrinsic relation for main sequence stars is plotted as a solid line, with the same relation reddened by E$_{B-V} = 0.38$ and E$_{B-V} = 0.70$ shown by dotted lines. The reddening relations for stars of spectral type B6.5 V and A2 V are shown as dashed lines.}
\label{fig3}
\end{figure}

\section{Berkeley 58}
Fig. 3 is a {\it UBV} colour-colour diagram for the field of Berkeley 58 surveyed in this study, as constructed from the data of Tables 1, 3, and 4. The phase-averaged data for CG Cas are from \citet{be07}. A reddened sequence of B and A-type cluster members can be detected in the data, but a cluster reddening of ${\rm E}_{B-V} \simeq 0.7$ places them in a section of the colour-colour diagram where they can be confused photometrically with unreddened, foreground, G-type stars. For that reason it becomes essential to make the process of photometric identification of likely spectral classes for individual stars as reliable as possible, through the use of a well-established interstellar extinction relation. The spectral types obtained for six of the B-type, photoelectrically-observed, cluster stars imply a reddening law for Berkeley 58 described by ${\rm E}_{U-B}/{\rm E}_{B-V} = 0.75$, along with a small curvature term \citep{tu89}, identical to the reddening slope found previously for star clusters spatially adjacent to Berkeley 58 \citep{tu76b}. Berkeley 58 stars were therefore dereddened with such a relationship, except for late-type stars where a steeper relationship was adopted, dependent upon the likely intrinsic colours of the stars.

\begin{figure}
\begin{center}
\includegraphics[width=6cm]{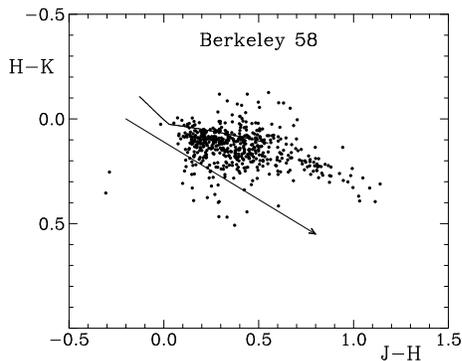}
\end{center}
\caption{A 2MASS colour-colour diagram, {\it H--K} versus {\it J--H}, for stars examined in the field of Berkeley 58, without regard to the uncertainties in the observations \citep{cu03}. The intrinsic relation for main sequence stars is plotted as a solid line, as derived from the observed colours of standard stars and stars in clusters of uniform reddening. The direction of reddening in the 2MASS system is indicated.}
\label{fig4}
\end{figure}

The Fig. 3 data indicate an absence of any unreddened O, B, or A-type stars in the observed sample. That feature is confirmed by available 2MASS data for the observed stars \citep{cu03}, which are depicted in the {\it JHK} colour-colour diagram of Fig. 4. An intrinsic relation for main-sequence stars in the 2MASS system was constructed from 2MASS observations of unreddened standard stars and stars in open clusters of uniform reddening \citep[e.g.,][]{tu96b}, adjusted with a reddening slope ${\rm E}_{H-K}/{\rm E}_{J-H} = 0.55$, as derived from reddened stars of known spectral type. The number of cluster stars with {\it U}-band observations is a small fraction of the total sample, so Fig. 4 contains many more stars than Fig. 3. The selection of 2MASS data was also not restricted according to the magnitude of cited uncertainties in the data, so several points in Fig. 4 display unusually large scatter. It seems clear, however, that the sample of cluster stars surveyed consists mainly of stars reddened by ${\rm E}_{J-H} \ge 0.1$, which corresponds to ${\rm E}_{B-V} \ge 0.36$.

The correleation of reddening with distance towards Berkeley 58 was established from the available {\it UBV} photometry by dereddening the colours for individual stars in conjunction with a copy of the POSS field on which derived colour excesses E$_{B-V}$ were recorded as they were obtained, with multiple solutions resolved by reference to the reddenings for spatially adjacent stars as well as by the reddenings derived for the stars from their 2MASS colours (Fig. 4). In most cases the smaller {\it JHK} reddening of stars relative to those obtained from {\it UBV} colours was sufficient to resolve questions about likely intrinsic colours for the stars, but there were a number of ambiguous cases where the data from the two surveys yielded disparate solutions, {\it e.g.} 2MASS colours implying an early spectral type and {\it UBV} colours implying a late spectral type. Such cases were unimportant in the final analysis, but are curious nevertheless.

\begin{figure}
\begin{center}
\includegraphics[width=6cm]{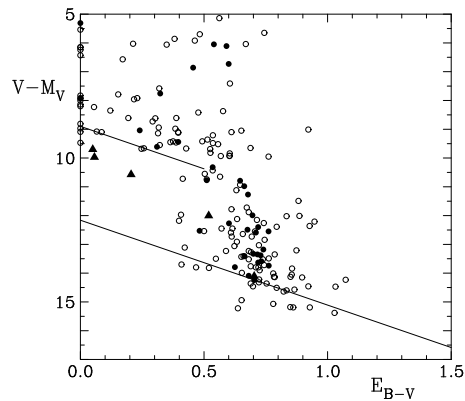}
\end{center}
\caption{A variable-extinction diagram for observed Berkeley 58 stars, with symbols as in Fig. 3. Reddening relations of slope $R=A_V/{\rm E}_{B-V}=2.95$ are shown corresponding to distances of $d\simeq 600$ pc ($V_0-M_V=8.9$) and $d\simeq 2700$ pc ($V_0-M_V=12.16$).}
\label{fig5}
\end{figure}

Distance moduli were calculated for individual stars by adoption of zero-age main sequence (ZAMS) values of $M_V$ \citep{tu76a,tu79}, so the values systematically underestimate {\it V--M}$_V$ for unresolved binaries and evolved stars. The resulting scatter in the variable-extinction diagram of Fig. 5 therefore contains a systematic component towards small values of {\it V--M}$_V$. Within such constraints, it is possible to discern certain trends in the data, such as the lack of any significant reddening out to distances of $\sim600$ pc ($V_0-M_V=8.9$), with a reddening of E$_{B-V} \ge 0.4$ beyond that to distances of $\sim2700$ pc ($V_0-M_V=12.16$) or more. At the Galactic location of CG Cas ($l = 116^{\circ}.845$, $b = -1^{\circ}.315$), a more encompassing survey by \citet{nk80} implies a similar trend, with the reddening beginning at distances of $\sim400-900$ pc. Apparently the main extinction for stars in the direction of Berkeley 58 occurs near the far side of the local spiral arm feature.

But the picture is not that simple. When the derived reddenings are compared star-for-star in the field of Berkeley 58, there are no obvious trends with spatial location, and trends with distance are difficult to establish without highly accurate luminosities for the observed stars. It can be surmised that there is additional reddening occurring on the near side of the Perseus spiral arm, given the nature of the scatter in the colour excesses. Likely members of Berkeley 58 generally have reddenings of E$_{B-V} \simeq 0.70$, with larger values possibly arising from circumstellar extinction, particularly for late B-type stars where rapid rotation is common \citep[e.g.,][]{tu93,tu96a}. An identical feature is observed in the adjacent cluster NGC 7790 \citep{ta88}. A lower envelope trend for the reddened stars in Fig. 5 implies a ratio of total-to-selective extinction for the field of $R=A_V/{\rm E}_{B-V}=2.95\pm0.30$ from least squares and non-parametric analyses. The value is consistent with previous studies of clusters in this region of the Galaxy \citep{tu76b}, as well as with a value of $R\simeq2.95$ expected for local extinction described by a reddening slope of 0.75 \citep{tu96a}. For subsequent calculations a value of $R=2.95$ was adopted, the exact choice affecting estimates of distance but not the derived luminosity for CG Cas as a cluster member.

An observational colour-magnitude diagram for the sampled atars is presented in Fig. 6, with a ZAMS plotted for {\it V--M}$_V$ = 14.29, the apparent distance modulus at E$_{B-V}= 0.70$ for points on the lower relation of Fig. 5. Such parameters provide a reasonable fit to the data, but there remain anomalies requiring further examination. For example, Fig. 6 contains reddened B-type stars more luminous than the turnoff magnitude for a cluster containing CG Cas, a point also indicated in Fig. 3, where dashed relations indicate reddening lines for B6.5 V and A2 V stars, the former corresponding to the expected turnoff color [$(B-V)_0=-0.13$] for stars associated with a 4$^{\rm d}$.37 Cepheid \citep{tu96c}. Clearly the field contains a number of stars younger than the expected evolutionary age of CG Cas.

\begin{figure}
\begin{center}
\includegraphics[width=6cm]{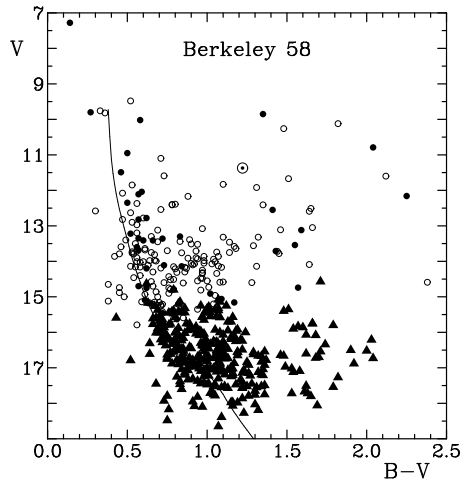}
\end{center}
\caption{A colour-magnitude diagram for Berkeley 58 from all observations: photoelectric (filled circles), photographic (open circles), and CCD (triangles) data. CG Cas is the circled point. The ZAMS is depicted for E$_{B-V}= 0.70$ and {\it V--M}$_V$ = 14.29.}
\label{fig6}
\end{figure}

\setcounter{table}{4}
\begin{table}
\caption[]{Radial Velocity Data for Berkeley 58 Stars.}
\label{tab5}
\begin{center}
\begin{tabular}{lccc}
\hline
Star &HJD &V$_{\rm R}$ &Adopted V$_{\rm R}$ \\
& &(km s$^{-1}$) &(km s$^{-1}$) \\
\hline
CG Cas &2445906.955 &$-74.5 \pm 3.8$ & \\
&2445908.942 &$-78.3 \pm1.4$ & \\
&2445909.944 &$-98.4 \pm3.0$ & \\
&2445910.933 &$-84.8 \pm1.3$ & \\
&2445911.935 &$-73.0 \pm1.8$ & \\
&2445912.923 &$-63.9 \pm1.7$ & \\
&2446326.874 &$-72.4 \pm2.3$ & \\
&2446327.907 &$-65.5 \pm3.7$ & \\
&2446328.910 &$-100.1 \pm1.2$ & \\
&2446330.890 &$-70.5 \pm1.2$ & \\
&2446331.881 &$-60.5 \pm2.3$ &$-78.8$ \\
\\
6 &2445908.961 &$-13.3 \pm3.5$ & \\
&2446326.940 &$-88.3 \pm4.1$ & \\
&2446327.955 &$-47.7 \pm6.5$ &$-52.3$ \\
\\
7 &2445909.963 &$-57.6 \pm5.3$ & \\
&2446327.942 &$-61.3 \pm2.9$ & \\
&2446331.020 &$-68.7 \pm4.2$ &$-62.7$ \\
\\
11 &2445910.956 &$-80.7 \pm1.2$ & \\
&2446330.919 &$-70.9 \pm5.1$ & \\
&2446331.909 &$-70.4 \pm3.3$ &$-79.1$ \\
\\
18 &2445911.760 &$-82.1 \pm10.1$ & \\
&2446326.914 &$-81.6 \pm3.3$ &$-81.6$ \\
\\
23 &2446328.955 &$-77.8 \pm13.0$ &$-77.8$ \\
\\
24 &2446330.972 &$-69.8 \pm8.5$ &$-69.8$ \\
\\
& &Cluster Mean = &$-79.4 \pm1.0$ \\
\hline
\end{tabular}
\end{center}
\end{table}

Such complications may be endemic to the field of both Berkeley 58 and NGC 7790, where the line of sight crosses the interarm region between the Sun and portions of the local spiral feature, then intercepts the Perseus spiral arm with a marked increase in space density for young B-type stars and young-to-intermediate age star clusters. The separation of spiral arm stars from cluster members is difficult but achievable, since the radial velocities for CG Cas and Berkeley 58 stars listed in Table 5 imply a conspicuous velocity difference between the cluster and spiral arm stars. The anomalously young B stars noted above are objects like stars 6 (V654 Cas), 7, and possibly 24, which have systematically more positive velocities than likely cluster members: stars 11, 18, and 23, which have radial velocities close to the systemic velocity of CG Cas \citep[see Fig. 7, which includes radial velocity measurements from][]{jo37,me91,go98}. Except for star 11, which may be anomalous, stars with radial velocities close to that of CG Cas also have spectral types near the expected B6.5 V turnoff. Unfortunately it is not possible to identify fainter cluster members by the same technique, given the bright limit for the present radial velocity survey. Follow-up observations would be useful in that regard.

\begin{figure}
\begin{center}
\includegraphics[width=6cm]{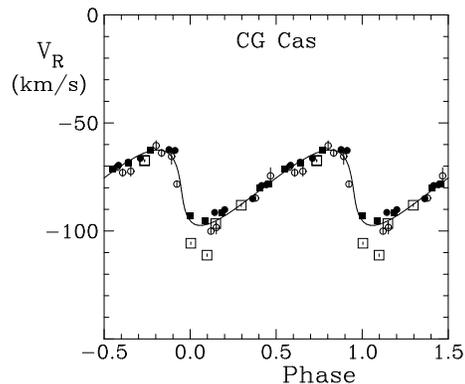}
\end{center}
\caption{The radial velocity variations of CG Cas, with cited uncertainties, as measured in this paper (open circles), from \citet{me91} and \citet{go98} (filled circles), and from \citet{jo37} (open squares). The curve is a simple spectroscopic binary solution to the data from the first three data sets, the data from \citet{jo37} exhibiting systematic deviations near velocity minimum.}
\label{fig7}
\end{figure}

The complications arising from contamination of the cluster field by young stars in the Perseus arm and likely circumstellar reddening for late B-type members were addressed by identifying unaffected cluster stars from their reddenings, which are close to E$_{B-V}{\rm (B0)} = 0.70$. The field of the CCD survey near the cluster centre was found to exhibit a mean reddening of E$_{B-V}{\rm (B0)} = 0.697 \pm0.025$, that for the region of CG Cas a mean reddening of E$_{B-V}{\rm (B0)} = 0.685 \pm0.022$. Stars with full {\it UBV} data were identified as likely cluster members on the basis of reddenings comparable to or larger than those values, while stars near the cluster centre lacking {\it U}-band data were assumed to have B0 star colour excesses as above, but intrinsic colours adjusted for the spectral type dependence of reddening \citep[see][]{fe63}. A-type dwarfs can suffer complications arising from the effects of rotation on their stellar continua and {\it UBV} colours \citep{te06b}, so the adoption of space reddenings for such stars may circumvent potential biases introduced by dereddening their colours to the intrinsic relation for zero-age zero-rotation main sequence stars. The resulting reddening-corrected colour-magnitude diagram for the cluster is plotted in Fig. 8 for 145 likely members, along with CG Cas and its light variations and star 5, which is considered to be a potential K giant member. The reddening for CG Cas corrected for its colour is E$_{B-V} = 0.64 \pm0.02$. A photometric reddening could be obtained from the {\it BVI$_c$} observations of \citet{he96} \citep[see][]{la07}, but a field reddening was adopted as a precaution against potential bias towards large-amplitude Cepheids lying near the centre of the instability strip (unnecessary in the present case, as it turns out). 

The distance to Berkeley 58 is established by 40 of its A-type ZAMS members, which yield a value of $V_0 - {\rm M}_V = 12.40 \pm 0.12$ s.d., corresponding to a distance of $3026 \pm166$ pc. Except for star 11, which is conceivably a rapid rotator observed nearly pole-on, the bluest cluster stars correspond to spectral type B6 with ({\it B--V})$_0 = -0.16$. A comparison with stellary evolutionary models \citep{me93} implies a cluster age of $10 \pm 1 \times 10^7$ years ($\log \tau = 8.0 \pm0.05$). The corresponding mass of cluster stars falling at the tip of the main-sequence red turnoff (RTO) is $5.4 M_{\sun}$ \citep{me93}.

\begin{figure}
\begin{center}
\includegraphics[width=6cm]{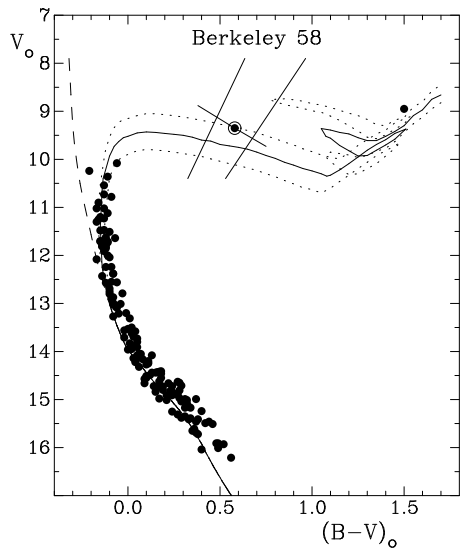}
\end{center}
\caption{A reddening free colour-magnitude diagram for Berkeley 58. The dashed curve represents the ZAMS for $V_0-M_V=12.40$, and the solid line with dotted lines on either side represents an isochrone from \citet{me93} for $\log \tau = 8.0 \pm0.1$. The range of light variations for CG Cas are depicted, as are the observational boundaries for the Cepheid instability strip. The red object on the evolved giant sequence is star 5.}
\label{fig8}
\end{figure}

\section{CG Cassiopeiae}
The systemic radial velocity of CG Cas (Table 5) is a close match to the mean velocity of Berkeley 58 derived from likely cluster members 11, 18, and 23, and the evolutionary age of the cluster closely matches what is predicted for the pulsation period of the Cepheid \citep{tu96c}. The luminosity of CG Cas as a likely member of Berkeley 58 is $\langle M_V \rangle = -3.06 \pm0.12$, which matches a value of $\langle M_V \rangle = -3.04$ predicted with a Cepheid period-radius relation and the inferred effective temperature of CG Cas ($\log T_{\rm eff} = 3.775$) from its derived intrinsic colour \citep{tb02}. The case for membership of CG Cas in Berkeley 58 is very strong.

The exact evolutionary status of CG Cas can be established from the direction and rate of its period changes \citep{te06a}, in conjunction with its large blue light amplitude of $\Delta B = 1.22$ \citep{be07}. The period changes for CG Cas were established here from examination of archival photographic plates in the Harvard and Sternberg collections, as well as from an analysis of new and existing photometry for the star. A working ephemeris for CG Cas based upon the available data was:
\begin{displaymath}
{\rm JD}_{\rm max} = 2432436.94 + 4.3656292 \: E ,
\end{displaymath}
where $E$ is the number of elapsed cycles. An extensive analysis of all available observations produced the data summarized in Table 6, which lists the results for different epochs, the type of data analyzed (PG = photographic, VIS = visual telescopic observations, B = photoelectric {\it B}, and V = photoelectric {\it V}), the number of observations used to establish the times of light maximum, and the source of the observations, in addition to the temporal parameters. The data are plotted in Fig. 9.

A regression analysis of the O--C data of Table 6 produced a parabolic solution for the ephemeris defined by:
\begin{displaymath}
{\rm JD}_{\rm max} = 2432436.9493(\pm0.0080)
\end{displaymath}
\begin{displaymath}
+ 4.3656289(\pm0.0000024) \: E + 1.1757(\pm0.0983) \times 10^{-7} \:E^2 ,
\end{displaymath}
which is plotted in Fig. 8. The parabolic trend corresponds to a period increase of $+0.170 \pm 0.014$ s yr$^{-1}$ ($ \log{\dot{P}}= -0.770 \pm 0.036$), a value typical of Cepheids lying slightly blueward of the centre of the instability strip and in the third crossing. The location of CG Cas in Fig. 8 relative to the observational boundaries of the Cepheid instability strip \citep{te06b} is consistent with that conclusion, although the stellar evolutionary models seem to require adjustments (metallicity, mixing of surface layers?) to match the observations.

\begin{figure}
\begin{center}
\includegraphics[width=6cm]{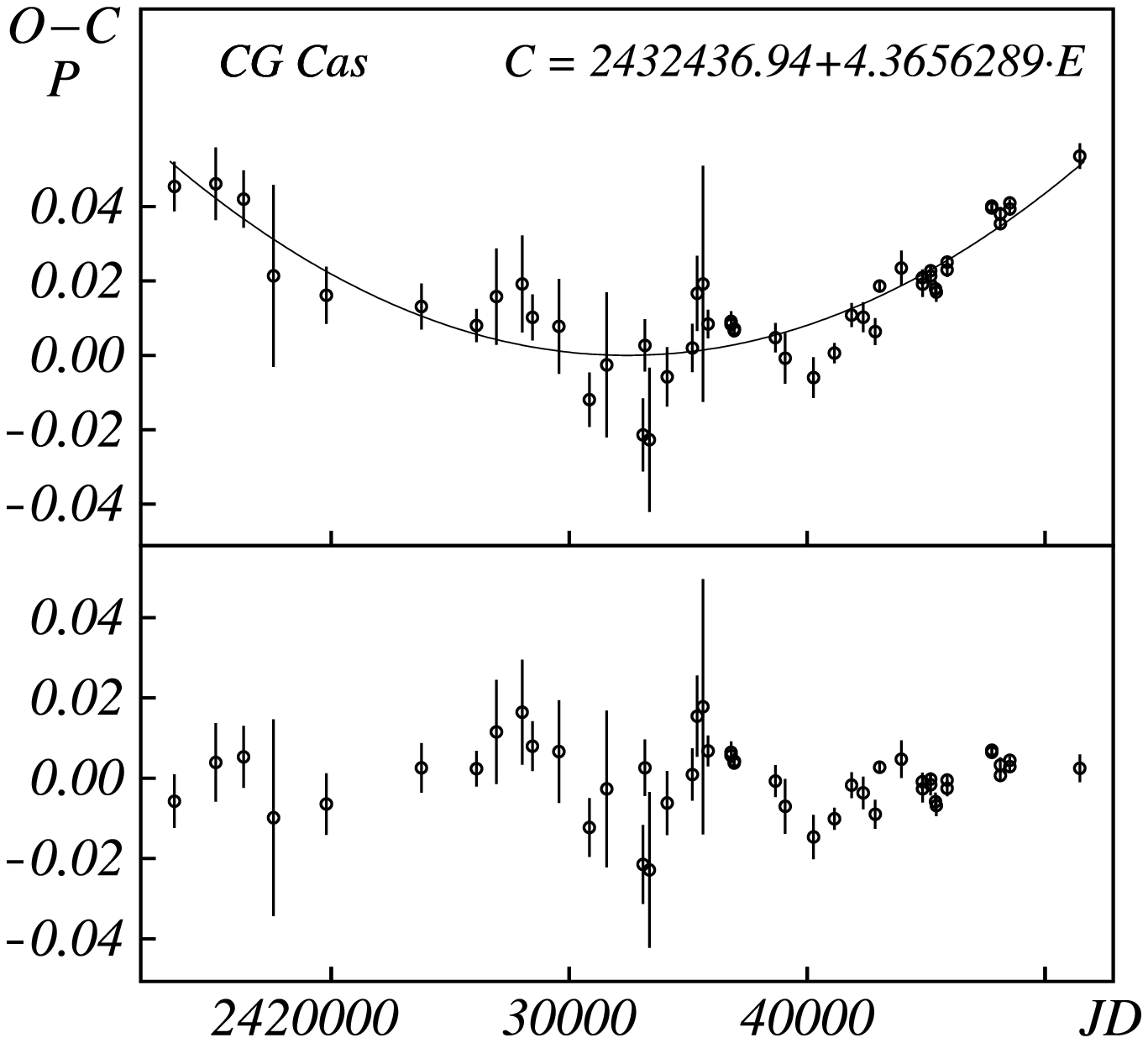}
\end{center}
\caption{The differences between observed (O) and computed (C) times of light maximum for CG Cas, computed in units of pulsation phase. The upper diagram shows the actual O--C variations with their uncertainties, the lower diagram the residuals from the calculated parabolic evolutionary trend.}
\label{fig9}
\end{figure}

\setcounter{table}{5}
\begin{table*}
\begin{minipage}{13cm}
\caption[]{Times of Maximum Light for CG Cas.}
\label{tab6}
\begin{tabular}{ccccccl}
\hline
HJD$_{\rm max}$ &$\pm \sigma$ &Band &Epoch &O--C &Observations &Reference \\
& & &(E) &(phase) &(n) & \\
\hline
2413407.3442 &0.0292 &PG &--4359 &+0.1714 &55 &This paper (Harvard) \\
2415144.8677 &0.0428 &PG &--3961 &+0.1746 &7 &This paper (SAI )\\
2416314.8382 &0.0338 &PG &--3693 &+0.1566 &72 &This paper (Harvard) \\
2417572.0492 &0.1070 &PG &--3405 &+0.0664 &11 &This paper (SAI) \\
2419794.1315 &0.0336 &PG &--2896 &+0.0436 &63 &This paper (Harvard) \\
2423788.6688 &0.0271 &PG &--1981 &+0.0304 &98  &This paper (Harvard) \\
2426102.4299 &0.0196 &PG &--1451 &+0.0082 &128 &This paper (Harvard) \\
2426940.6916 &0.0567 &VIS &--1259 &+0.0691 &46 &\citet{la33} \\
2428023.3553 &0.0571 &PG &--1011 &+0.0569 &19 &This paper (SAI) \\
2428455.5134 &0.0271 &PG &--912 &+0.0177 &92 &This paper (Harvard) \\
2429568.7382 &0.0559 &PG &--657 &+0.0071 &28 &This paper (SAI) \\
2430847.7814 &0.0321 &PG &--364 &--0.0790 &81 &This paper (Harvard) \\
2431576.8823 &0.0854 &PG &--197 &--0.0381 &17 &\citet{er61} \\
2433100.4046 &0.0430 &PG &+152 &--0.1203 &59 &This paper (Harvard) \\
2433183.4566 &0.0308 &PG &+171 &--0.0152 &37 &This paper (SAI) \\
2433371.0678 &0.0848 &PG &+214 &--0.1261 &23 &\citet{er61} \\
2434117.6643 &0.0350 &PG &+385 &--0.0521 &25 &This paper (SAI) \\
2435174.1804 &0.0285 &PG &+627 &--0.0182 &74 &This paper (SAI) \\
2435379.4291 &0.0443 &PG &+674 &+0.0459 &10 &\citet{ro59} \\
2435619.5498 &0.1388 &PG &+729 &+0.0570 &19 &\citet{er61} \\
2435837.7841 &0.0168 &PG &+779 &+0.0099 &18 &\citet{zs59} \\
2436802.5876 &0.0070 &B &+1000 &+0.0094 &13 &\citet{oo60} \\
2436802.6183 &0.0119 &V &+1000 &+0.0401 &15 &\citet{oo60} \\
2436933.5492 &0.0054 &B &+1030 &+0.0021 &22 &\citet{ba62} \\
2436937.9440 &0.0085 &V &+1031 &+0.0313 &23 &\citet{ba62} \\
2438666.6957 &0.0174 &PG &+1427 &--0.0061 &41 &This paper (SAI) \\ 
2439077.0406 &0.0299 &PG &+1521 &--0.0303 &16 &This paper (SAI) \\
2440268.8346 &0.0241 &PG &+1794 &--0.0530 &24 &This paper (SAI) \\
2441146.3548 &0.0121 &PG &+1995 &--0.0242 &95 &This paper (SAI) \\
2441866.7282 &0.0142 &PG &+2160 &+0.0204 &55 &This paper (SAI) \\
2442355.6761 &0.0178 &PG &+2272 &+0.0178 &47 &This paper (SAI) \\
2442862.0722 &0.0159 &PG &+2388 &+0.0010 &74 &This paper (SAI) \\
2443045.5091 &0.0058 &V &+2430 &+0.0815 &71 &\citet{ch82} \\
2443957.9197 &0.0206 &PG &+2639 &+0.0756 &25 &This paper (SAI) \\
2444844.1310 &0.0099 &B &+2842 &+0.0643 &9 &\citet{be86} \\
2444852.8817 &0.0150 &V &+2844 &+0.0837 &11 &\citet{be86} \\
2445189.0177 &0.0117 &B &+2921 &+0.0663 &8 &\citet{be86} \\
2445189.0509 &0.0074 &V &+2921 &+0.0995 &8 &\citet{be86} \\
2445394.1872 &0.0098 &B &+2968 &+0.0512 &14 &This paper \\
2445429.1355 &0.0115 &V &+2976 &+0.0745 &15 &This paper \\
2445883.1690 &0.0061 &B &+3080 &+0.0826 &8 &\citet{be86} \\
2445883.1870 &0.0086 &V &+3080 &+0.1006 &8 &\citet{be86} \\
2447760.4530 &0.0042 &B &+3510 &+0.1461 &39 &\citet{be92a} \\
2447760.4823 &0.0059 &V &+3510 &+0.1754 &39 &\citet{be92a} \\
2448118.4162 &0.0060 &B &+3592 &+0.1277 &18 &\citet{be92b} \\
2448118.4546 &0.0085 &V &+3592 &+0.1661 &18 &\citet{be92b} \\
2448515.7127 &0.0043 &B &+3683 &+0.1520 &20 &\citet{be92c} \\
2448515.7328 &0.0052 &V &+3683 &+0.1721 &20 &\citet{be92c} \\
2451458.2287 &0.0152 &V &+4357 &+0.2341 &27 &\citet{wo04} \\
\hline
\end{tabular}
\end{minipage}
\end{table*}

\section{Discussion}
The case for potential membership of the Cepheid CG Cas in the sparse open cluster Berkeley 58 has been studied using photometric (pe, pg, CCD) observations, spectroscopy (V$_{\rm R}$, spectral types), star counts, and O--C data for the Cepheid. The cluster Berkeley 58 is particularly difficult to separate from the young stars of the Perseus spiral arm, which raises concerns about future studies of distant open cluster calibrators for the Cepheid PL relation. Careful analysis of the available data leads to a cluster reddening of E$_{B-V}{\rm (B0)} = 0.70$, a distance of $3.03 \pm0.17$ kpc, and an age of $10 \pm 1 \times 10^7$ years. CG Cas is a likely member on the basis of radial velocity, location outside the cluster nucleus within the cluster coronal region, evolutionary status indicated by its period changes and light amplitude, and implied luminosity. It becomes an important Cepheid calibrator lying near the centre of the instability strip.

It may seem unusual that many potential Cepheid calibrators lie in cluster coronae rather than cluster nuclear regions \citep{tu85}, but a possible explanation relates to two dynamical lines of evidence. First, massive cluster members lie preferentially in outer regions of clusters \citep{bu78}, possibly because of how proto-cluster interstellar clouds fragment into proto-stars. Second, as indicated by colour-magnitude diagrams for NGC 654 \citep{st80} and other young clusters \citep{tu96b}, cluster nuclear regions tend to be dominated by rapidly rotating stars, possibly the result of merged binary systems, and other close binaries, in which case potential Cepheid progenitors are less likely to evolve to the dimensions typical of pulsating variables because of restrictions on their dimensions engendered by potential physical companions. The case of CG Cas in Berkeley 58 appears to be yet another example of the effect.

\subsection*{ACKNOWLEDGEMENTS}
The present study was supported by research funding awarded through the Natural Sciences and Engineering Research Council of Canada (NSERC), through the Small Research Grants program of the American Astronomical Society, through the Russian Foundation for Basic Research (RFBR), and through the program of Support for Leading Scientific Schools of Russia. We are endebted to Ron Lyons for scanning the radial velocity plates used in this study, and to the director of Harvard College Observatory for access to the plate stacks.

\setcounter{table}{2}
\begin{table*}
\begin{minipage}{17cm}
\subsection*{APPENDIX}
\caption{Photographic {\it UBV} Data for Stars in Berkeley 58. }
\label{tab3}
\begin{tabular}{ccccccccccccc}
\hline
Star &RA(2000) &DEC(2000) &$V$ &$B-V$ &$U-B$ & &Star &RA(2000) &DEC(2000) &$V$ &$B-V$ &$U-B$ \\
\hline
101 &23 59 36.09 &+60 49 01.9 &9.48 &0.52 &$-0.05$ & &179 &00 01 21.27 &+60 53 30.5 &14.01 &0.60 &+0.47 \\
102 &00 00 13.86 &+61 04 47.4 &9.76 &0.33 &+0.15 & &180 &23 59 56.23 &+60 56 25.0 &14.03 &0.58 &+0.08 \\
103 &00 01 16.38 &+60 49 18.1 &9.82 &0.36 &$-0.28$ & &181 &23 58 53.10 &+60 57 05.6 &14.05 &0.98 &+0.93 \\
104 &23 59 24.28 &+60 48 18.3 &10.12 &1.82 &+0.74 & &182 &23 59 10.64 &+60 51 55.7 &14.06 &0.90 &+0.25 \\
105 &23 59 35.71 &+60 47 49.6 &10.26 &1.48 &+0.84 & &183 &23 59 17.01 &+60 51 12.5 &14.07 &1.03 &+0.19 \\
106 &00 00 03.42 &+60 50 36.1 &11.10 &0.71 &$-0.05$ & &184 &23 59 47.40 &+60 48 27.0 &14.07 &0.92 &+0.30 \\
107 &23 59 36.15 &+60 47 33.1 &11.59 &0.73 &+0.13 & &185 &23 59 50.39 &+60 58 39.3 &14.09 &1.64 &... \\
108 &23 59 42.73 &+60 47 16.1 &11.60 &2.12 &+2.67 & &186 &00 00 49.32 &+60 48 06.5 &14.10 &0.91 &+0.08 \\
109 &23 59 07.11 &+60 55 06.8 &11.67 &1.51 &+1.09 & &187 &00 00 24.36 &+60 55 35.8 &14.11 &0.63 &+0.09 \\
110 &00 01 05.72 &+60 51 17.7 &11.83 &1.10 &+0.67 & &188 &23 59 49.10 &+60 50 32.7 &14.13 &1.07 &+0.35 \\
111 &00 00 11.91 &+60 50 28.6 &11.85 &0.53 &+0.17 & &189 &23 58 57.51 &+60 58 24.8 &14.14 &1.31 &+0.85 \\
112 &00 01 20.84 &+60 52 34.9 &11.92 &1.31 &+0.95 & &190 &23 59 22.80 &+60 57 04.3 &14.14 &0.52 &$-0.12$ \\
113 &00 01 21.00 &+61 00 48.4 &12.08 &0.47 &+0.31 & &191 &23 59 05.78 &+61 01 43.6 &14.15 &0.79 &+0.70 \\
114 &00 01 10.16 &+61 03 50.0 &12.17 &0.88 &+0.34 & &192 &23 59 46.99 &+61 03 27.0 &14.16 &0.58 &+0.38 \\
115 &00 01 20.63 &+60 55 32.1 &12.37 &0.56 &+0.12 & &193 &00 00 55.95 &+61 03 02.4 &14.16 &0.55 &+0.19 \\
116 &00 00 07.17 &+60 48 48.4 &12.39 &0.80 &+0.52 & &194 &00 00 04.13 &+60 51 51.7 &14.17 &0.86 &+0.54 \\
117 &23 58 53.91 &+60 56 37.2 &12.40 &0.78 &+0.52 & &195 &00 00 37.31 &+60 46 36.1 &14.17 &1.09 &+0.43 \\
118 &23 59 10.07 &+60 55 48.6 &12.41 &0.78 &+0.52 & &196 &23 59 37.23 &+61 01 47.2 &14.19 &0.71 &$-0.04$ \\
119 &23 59 40.81 &+60 51 12.4 &12.41 &1.35 &+1.05 & &197 &00 01 14.97 &+60 54 01.9 &14.20 &0.87 &+0.26 \\
120 &00 00 24.15 &+61 05 54.4 &12.51 &1.65 &+1.73 & &198 &23 59 07.54 &+60 59 40.0 &14.21 &0.80 &$-0.01$ \\
121 &00 00 16.78 &+60 52 39.3 &12.54 &0.71 &+0.22 & &199 &23 59 49.51 &+60 59 23.1 &14.25 &0.97 &+0.85 \\
122 &00 01 05.00 &+60 50 58.3 &12.58 &0.30 &$-0.02$ & &200 &00 00 26.46 &+60 50 28.7 &14.30 &0.66 &+0.52 \\
123 &23 59 11.61 &+61 02 04.7 &12.59 &1.64 &+0.71 & &201 &00 00 51.81 &+60 46 54.6 &14.30 &0.86 &+0.22 \\
124 &23 59 43.07 &+61 03 17.7 &12.63 &0.52 &+0.20 & &202 &00 00 23.33 &+60 51 42.1 &14.31 &0.81 &+0.21 \\
125 &23 59 06.63 &+60 53 17.9 &12.78 &0.60 &+0.33 & &203 &23 59 34.32 &+60 59 24.9 &14.32 &0.97 &+0.31 \\
126 &00 01 05.84 &+60 59 50.1 &12.82 &0.47 &$-0.01$ & &204 &23 59 17.67 &+60 54 45.5 &14.34 &1.06 &+0.33 \\
127 &00 01 33.08 &+60 53 08.0 &12.90 &0.53 &+0.41 & &205 &00 00 11.83 &+61 05 55.5 &14.34 &0.97 &+0.41 \\
128 &00 00 57.98 &+61 04 02.5 &12.97 &0.56 &$-0.20$ & &206 &23 59 41.42 &+60 51 28.7 &14.40 &0.58 &+0.40 \\
129 &00 00 25.44 &+60 59 52.4 &13.05 &1.66 &+1.57 & &207 &00 00 25.66 &+60 50 43.7 &14.43 &0.76 &+0.18 \\
130 &23 59 03.83 &+60 51 31.8 &13.08 &0.97 &+0.23 & &208 &23 59 42.89 &+61 02 29.1 &14.46 &0.54 &+0.03 \\
131 &23 58 54.24 &+60 54 15.1 &13.11 &1.46 &+1.48 & &209 &00 00 58.82 &+60 55 01.6 &14.47 &0.96 &+0.59 \\
132 &00 00 26.72 &+60 59 55.4 &13.12 &0.69 &+0.29 & &210 &00 00 00.99 &+61 00 51.4 &14.48 &0.97 &+0.78 \\
133 &23 59 19.28 &+60 50 11.7 &13.20 &0.56 &$-0.04$ & &211 &00 01 25.15 &+61 02 11.0 &14.51 &0.79 &+0.56 \\
134 &23 59 24.57 &+60 55 27.9 &13.30 &0.65 &+0.40 & &212 &23 59 23.62 &+60 59 41.4 &14.52 &0.77 &+0.36 \\
135 &23 59 59.65 &+61 04 09.7 &13.32 &1.32 &+1.21 & &213 &00 00 26.84 &+60 49 40.7 &14.55 &1.00 &+0.51 \\
136 &00 00 27.04 &+60 46 43.8 &13.32 &1.10 &+0.22 & &214 &00 01 21.95 &+61 02 29.5 &14.55 &0.94 &+0.62 \\
137 &00 00 28.97 &+60 47 58.6 &13.39 &0.89 &+0.24 & &215 &00 00 25.83 &+60 55 38.0 &14.57 &0.65 &+0.28 \\
138 &00 00 21.24 &+60 51 10.6 &13.40 &0.49 &+0.20 & &216 &23 59 06.22 &+60 53 57.6 &14.58 &1.10 &+0.46 \\
139 &00 01 16.05 &+61 02 45.5 &13.44 &0.83 &+0.55 & &217 &00 01 18.21 &+61 01 23.9 &14.59 &1.08 &+0.49 \\
140 &23 59 41.35 &+61 05 46.4 &13.45 &0.69 &+0.25 & &218 &00 00 16.19 &+60 53 17.5 &14.59 &2.38 &... \\
141 &23 58 50.90 &+60 54 37.2 &13.46 &1.14 &+0.16 & &219 &23 59 22.49 &+60 52 00.0 &14.60 &1.28 &+0.28 \\
142 &00 00 15.19 &+60 59 41.4 &13.50 &0.55 &+0.08 & &220 &00 00 01.21 &+60 57 39.2 &14.60 &0.73 &+0.47 \\
143 &00 01 38.02 &+60 57 11.3 &13.51 &0.90 &+0.24 & &221 &00 00 28.99 &+60 52 57.1 &14.61 &0.85 &+0.31 \\
144 &00 00 19.83 &+60 49 11.5 &13.52 &0.94 &+0.48 & &222 &00 00 53.50 &+60 55 22.0 &14.65 &0.38 &+0.08 \\
145 &23 59 48.55 &+60 46 45.3 &13.56 &1.29 &+0.38 & &223 &00 00 28.35 &+61 05 20.9 &14.70 &0.66 &$-0.01$ \\
146 &00 00 24.12 &+60 58 43.8 &13.58 &1.17 &+0.83 & &224 &23 59 58.59 &+60 54 09.2 &14.73 &1.01 &+0.71 \\
147 &00 01 14.18 &+60 56 32.2 &13.59 &0.46 &+0.03 & &225 &00 01 16.76 &+60 54 30.9 &14.74 &0.45 &$-0.17$ \\
148 &23 59 24.83 &+60 59 52.9 &13.64 &1.18 &... & &226 &00 00 18.03 &+60 51 36.5 &14.75 &0.76 &+0.37 \\
149 &00 00 07.41 &+61 00 24.1 &13.64 &0.61 &+0.23 & &227 &23 59 32.73 &+60 59 21.3 &14.79 &0.79 &+0.48 \\
150 &00 00 01.78 &+60 48 57.5 &13.68 &0.61 &+0.30 & &228 &00 00 10.51 &+60 58 01.8 &14.79 &0.77 &+0.43 \\
151 &23 58 55.65 &+60 52 55.3 &13.70 &0.75 &+0.27 & &229 &23 59 40.12 &+60 57 06.2 &14.81 &0.96 &+0.33 \\
152 &23 59 09.72 &+60 55 33.4 &13.71 &0.93 &+0.31 & &230 &00 00 20.17 &+60 55 54.3 &14.86 &0.68 &+0.28 \\
153 &23 59 52.24 &+60 56 50.2 &13.72 &0.56 &+0.13 & &231 &00 00 02.48 &+60 54 42.4 &14.88 &0.45 &+0.20 \\
154 &00 00 49.48 &+61 03 45.2 &13.72 &1.44 &+0.78 & &232 &23 59 49.71 &+60 53 31.9 &14.90 &0.76 &+0.39 \\
155 &00 01 38.95 &+60 57 40.8 &13.73 &0.53 &$-0.01$ & &233 &00 00 04.01 &+60 54 47.8 &14.93 &0.61 &+0.19 \\
156 &00 00 27.09 &+61 05 48.8 &13.74 &1.06 &+0.32 & &234 &00 00 33.02 &+60 55 51.9 &14.96 &1.05 &+0.44 \\
157 &23 59 34.31 &+60 49 44.5 &13.75 &1.14 &+0.47 & &235 &00 00 09.83 &+60 54 30.1 &15.00 &0.69 &+0.21 \\
158 &00 00 44.24 &+60 46 52.3 &13.75 &0.65 &+0.08 & &236 &00 00 44.08 &+61 05 04.7 &15.00 &0.48 &+0.49 \\
159 &00 00 31.53 &+60 46 32.0 &13.76 &0.84 &+0.34 & &237 &00 00 09.76 &+61 05 34.5 &15.12 &0.38 &+0.26 \\
160 &00 00 38.87 &+60 53 10.0 &13.77 &0.53 &+0.09 & &238 &23 59 42.48 &+60 56 18.8 &15.14 &0.59 &+0.36 \\
\hline
\end{tabular}
\end{minipage}
\end{table*}

\setcounter{table}{2}
\begin{table*}
\begin{minipage}{17cm}
\caption{Continued.}
\label{tab3}
\begin{tabular}{ccccccccccccc}
\hline
Star &RA(2000) &DEC(2000) &$V$ &$B-V$ &$U-B$ & &Star &RA(2000) &DEC(2000) &$V$ &$B-V$ &$U-B$ \\
\hline
161 &23 59 26.17 &+60 49 45.8 &13.78 &1.45 &+1.09 & &239 &00 00 04.71 &+60 57 45.6 &15.15 &0.67 &+0.44 \\
162 &23 59 41.19 &+61 04 51.7 &13.79 &0.54 &+0.27 & &240 &23 59 19.79 &+61 00 25.5 &15.16 &0.70 &+0.13 \\
163 &00 01 38.40 &+60 56 46.7 &13.79 &0.45 &+0.29 & &241 &23 59 39.28 &+60 57 14.7 &15.17 &0.64 &+0.31 \\
164 &00 01 13.36 &+61 01 33.3 &13.81 &0.67 &+0.05 & &242 &00 00 22.59 &+60 57 40.8 &15.17 &0.63 &+0.09 \\
165 &23 59 20.85 &+61 02 22.0 &13.86 &0.98 &+0.61 & &243 &23 59 53.74 &+60 57 08.8 &15.20 &1.04 &+0.39 \\
166 &00 00 05.82 &+60 50 35.8 &13.86 &0.42 &+0.00 & &244 &00 00 04.46 &+61 00 44.7 &15.21 &0.72 &+0.28 \\
167 &00 00 02.37 &+60 46 38.7 &13.86 &1.07 &+0.04 & &245 &00 00 06.29 &+60 54 44.8 &15.21 &0.66 &+0.29 \\
168 &23 59 04.31 &+61 01 44.6 &13.90 &0.93 &+0.84 & &246 &00 00 38.77 &+60 56 31.4 &15.21 &0.82 &+0.49 \\
169 &00 00 00.15 &+60 55 14.2 &13.90 &0.56 &+0.06 & &247 &23 59 50.61 &+60 55 25.6 &15.23 &0.71 &... \\
170 &23 59 03.70 &+61 01 50.1 &13.91 &0.88 &+0.87 & &248 &23 59 57.24 &+60 55 02.8 &15.24 &0.69 &+0.29 \\
171 &00 01 09.90 &+60 52 54.7 &13.93 &0.94 &+0.43 & &249 &23 59 55.11 &+60 53 44.8 &15.26 &0.71 &+0.31 \\
172 &00 01 08.86 &+60 58 34.1 &13.94 &1.04 &+0.66 & &250 &00 00 08.27 &+60 56 42.6 &15.35 &0.94 &+0.61 \\
173 &23 59 27.78 &+60 55 30.8 &13.97 &0.67 &+0.49 & &251 &00 00 14.46 &+60 57 47.6 &15.42 &0.77 &+0.39 \\
174 &00 00 13.79 &+61 01 04.7 &13.98 &0.98 &+0.48 & &252 &00 00 15.30 &+60 54 57.0 &15.42 &0.72 &... \\
175 &00 00 26.02 &+60 55 07.4 &13.98 &0.64 &+0.20 & &253 &00 00 31.15 &+61 04 07.7 &15.59 &0.85 &... \\
176 &00 01 10.45 &+60 58 30.7 &13.99 &0.68 &+0.42 & &254 &00 00 10.69 &+60 55 58.4 &15.62 &0.79 &... \\
177 &00 00 11.44 &+60 51 41.8 &14.01 &0.56 &+0.12 & &255 &00 00 06.19 &+60 56 32.6 &15.79 &0.56 &+0.23 \\
178 &00 01 16.45 &+60 58 25.5 &14.01 &1.05 &+0.21 & & & & & & & \\
\hline
\end{tabular}
\end{minipage}
\end{table*}

\setcounter{table}{3}
\begin{table*}
\begin{minipage}{17cm}
\caption{CCD {\it UBV} Data for Stars in the Nucleus of Berkeley 58.}
\label{tab4}
\begin{tabular}{ccccccccccccc}
\hline
Star &RA(2000) &DEC(2000) &$V$ &$B-V$ &$U-B$ & &Star &RA(2000) &DEC(2000) &$V$ &$B-V$ &$U-B$ \\
\hline
1001 &23 59 13.34 &+60 54 41.7 &16.66 &0.95 &... & &1192 &23 59 59.79 &+60 53 21.0 &16.89 &1.08 &... \\
1002 &23 59 13.58 &+60 55 25.5 &16.14 &0.73 &... & &1193 &00 00 00.31 &+60 54 14.4 &16.54 &1.36 &... \\
1003 &23 59 12.82 &+60 57 28.6 &15.57 &1.08 &... & &1194 &00 00 04.01 &+61 01 58.9 &16.48 &1.09 &... \\
1004 &23 59 12.99 &+60 56 53.1 &15.94 &0.72 &... & &1195 &00 00 01.08 &+60 55 09.3 &16.33 &1.02 &... \\
1005 &23 59 12.00 &+60 53 31.9 &17.75 &1.04 &... & &1196 &00 00 02.86 &+60 58 36.1 &15.55 &0.70 &... \\
1006 &23 59 14.92 &+61 00 43.2 &15.93 &1.05 &... & &1197 &00 00 04.28 &+61 01 21.8 &17.45 &0.68 &... \\
1007 &23 59 15.93 &+61 02 43.6 &17.47 &1.05 &... & &1199 &00 00 04.99 &+61 02 05.1 &17.21 &1.35 &... \\
1008 &23 59 16.20 &+60 53 18.5 &17.02 &0.95 &... & &1200 &00 00 03.25 &+60 57 25.0 &16.32 &1.07 &... \\
1009 &23 59 15.49 &+61 00 26.4 &16.92 &1.26 &... & &1201 &00 00 01.48 &+60 52 41.4 &16.07 &1.62 &... \\
1010 &23 59 16.63 &+61 01 54.6 &17.06 &1.13 &... & &1203 &00 00 01.53 &+60 52 25.7 &15.75 &0.87 &... \\
1011 &23 59 14.72 &+60 56 52.5 &17.15 &1.16 &... & &1205 &00 00 05.80 &+61 01 13.8 &16.77 &0.96 &... \\
1012 &23 59 14.95 &+60 57 04.1 &16.56 &1.09 &... & &1206 &00 00 02.37 &+60 53 20.1 &16.78 &0.91 &... \\
1013 &23 59 16.05 &+60 59 17.2 &14.57 &1.71 &... & &1207 &00 00 02.97 &+60 54 13.8 &15.84 &0.93 &... \\
1014 &23 59 13.40 &+60 52 36.2 &16.16 &0.85 &... & &1208 &00 00 02.45 &+60 52 53.0 &16.66 &1.11 &... \\
1015 &23 59 14.93 &+60 56 09.3 &16.38 &1.08 &... & &1209 &00 00 04.49 &+60 57 29.5 &16.99 &1.00 &... \\
1016 &23 59 13.64 &+60 53 08.8 &16.38 &1.00 &... & &1211 &00 00 05.98 &+60 59 42.6 &17.07 &0.88 &... \\
1017 &23 59 15.15 &+60 56 21.8 &16.61 &0.77 &... & &1213 &00 00 04.22 &+60 54 26.9 &16.61 &0.64 &... \\
1018 &23 59 15.09 &+60 56 05.4 &16.29 &0.99 &... & &1214 &00 00 04.64 &+60 55 16.8 &16.97 &0.91 &... \\
1019 &23 59 15.91 &+60 57 50.1 &18.08 &0.92 &... & &1215 &00 00 03.56 &+60 52 40.1 &15.33 &1.00 &... \\
1020 &23 59 17.35 &+61 00 48.8 &17.05 &1.04 &... & &1216 &00 00 08.29 &+61 02 39.0 &15.44 &0.86 &... \\
1021 &23 59 16.51 &+60 58 48.2 &17.44 &0.96 &... & &1218 &00 00 05.99 &+60 56 57.7 &17.20 &1.00 &... \\
1022 &23 59 16.58 &+60 58 23.4 &16.62 &1.29 &... & &1220 &00 00 07.02 &+60 58 12.1 &16.78 &0.81 &... \\
1023 &23 59 17.72 &+61 00 12.3 &17.82 &0.94 &... & &1222 &00 00 05.08 &+60 53 18.6 &16.86 &1.08 &... \\
1024 &23 59 15.99 &+60 55 24.2 &17.13 &0.86 &... & &1223 &00 00 05.00 &+60 53 09.0 &17.75 &1.36 &... \\
1025 &23 59 17.49 &+60 58 26.9 &15.97 &0.96 &... & &1224 &00 00 07.16 &+60 56 52.8 &17.97 &1.16 &... \\
1026 &23 59 17.43 &+60 57 55.6 &15.08 &1.07 &... & &1225 &00 00 05.25 &+60 52 21.2 &17.19 &1.34 &... \\
1027 &23 59 16.20 &+60 53 18.5 &16.77 &0.98 &... & &1227 &00 00 06.82 &+60 55 59.0 &16.19 &0.75 &... \\
1028 &23 59 17.45 &+60 55 32.8 &16.04 &0.98 &... & &1228 &00 00 08.19 &+60 58 11.0 &15.91 &0.90 &... \\
1029 &23 59 19.01 &+60 59 14.0 &17.23 &0.98 &... & &1229 &00 00 06.17 &+60 53 07.6 &17.72 &1.63 &... \\
1030 &23 59 17.76 &+60 55 55.7 &15.77 &0.66 &... & &1230 &00 00 06.64 &+60 53 58.4 &18.39 &1.09 &... \\
1031 &23 59 17.95 &+60 56 16.5 &16.04 &0.93 &... & &1231 &00 00 08.71 &+60 58 24.0 &16.19 &1.15 &... \\
1034 &23 59 20.75 &+61 01 52.0 &17.54 &1.79 &... & &1232 &00 00 07.31 &+60 55 02.8 &15.75 &1.66 &... \\
1035 &23 59 20.78 &+61 02 08.0 &15.90 &1.54 &... & &1233 &00 00 09.48 &+60 59 59.1 &16.64 &1.46 &... \\
1036 &23 59 19.06 &+60 57 40.1 &16.92 &1.36 &... & &1234 &00 00 09.94 &+61 01 12.3 &17.67 &1.03 &... \\
1037 &23 59 21.76 &+61 02 13.7 &16.38 &0.98 &... & &1235 &00 00 10.21 &+61 01 33.5 &15.35 &1.51 &... \\
1038 &23 59 21.29 &+61 00 29.4 &17.49 &1.36 &... & &1236 &00 00 07.39 &+60 54 47.3 &16.33 &0.78 &... \\
1039 &23 59 20.03 &+60 57 05.3 &17.09 &1.10 &... & &1238 &00 00 11.06 &+61 02 06.7 &17.04 &1.15 &... \\
1040 &23 59 19.65 &+60 54 46.4 &16.78 &1.37 &... & &1240 &00 00 08.12 &+60 54 44.7 &16.01 &0.70 &... \\
\hline
\end{tabular}
\end{minipage}
\end{table*}

\setcounter{table}{3}
\begin{table*}
\begin{minipage}{17cm}
\caption{CCD {\it UBV} Data for Stars in the Nucleus of Berkeley 58.}
\label{tab4}
\begin{tabular}{ccccccccccccc}
\hline
Star &RA(2000) &DEC(2000) &$V$ &$B-V$ &$U-B$ & &Star &RA(2000) &DEC(2000) &$V$ &$B-V$ &$U-B$ \\
\hline
1041 &23 59 18.71 &+60 52 29.0 &16.54 &1.71 &... & &1242 &00 00 07.23 &+60 52 10.7 &16.73 &2.04 &... \\
1042 &23 59 21.26 &+60 58 02.7 &17.11 &1.06 &... & &1243 &00 00 08.18 &+60 54 17.3 &16.51 &0.85 &... \\
1043 &23 59 19.98 &+60 54 26.0 &17.52 &1.04 &... & &1244 &00 00 12.27 &+61 03 25.6 &16.87 &1.17 &... \\
1044 &23 59 21.95 &+60 55 40.3 &16.45 &1.07 &... & &1246 &00 00 11.22 &+61 00 50.2 &17.57 &1.13 &... \\
1047 &23 59 21.29 &+61 00 29.4 &17.71 &1.02 &... & &1248 &00 00 09.33 &+60 55 04.1 &15.40 &0.99 &... \\
1048 &23 59 25.23 &+61 01 44.0 &18.17 &0.76 &... & &1249 &00 00 07.98 &+60 51 49.3 &17.45 &1.01 &... \\
1050 &23 59 25.80 &+61 00 40.3 &16.88 &0.78 &... & &1251 &00 00 11.50 &+60 58 32.1 &16.49 &1.22 &... \\
1051 &23 59 25.17 &+60 57 56.3 &16.88 &0.90 &... & &1254 &00 00 10.42 &+60 54 35.7 &16.55 &1.19 &... \\
1052 &23 59 23.33 &+60 53 29.9 &18.07 &1.16 &... & &1255 &00 00 13.91 &+61 02 16.6 &16.71 &1.12 &... \\
1053 &23 59 24.72 &+60 56 16.0 &15.59 &0.68 &... & &1256 &00 00 12.77 &+60 59 23.5 &17.12 &1.54 &... \\
1054 &23 59 25.52 &+60 58 09.7 &16.68 &0.99 &... & &1258 &00 00 13.64 &+61 00 37.4 &16.37 &0.74 &... \\
1056 &23 59 28.15 &+61 02 06.4 &15.35 &1.05 &... & &1260 &00 00 15.23 &+61 03 28.2 &16.50 &1.06 &... \\
1057 &23 59 28.38 &+61 02 40.0 &17.79 &1.26 &... & &1261 &00 00 11.36 &+60 53 43.7 &16.56 &1.19 &... \\
1058 &23 59 25.80 &+60 56 12.5 &16.11 &0.74 &... & &1262 &00 00 15.48 &+61 02 13.4 &17.22 &1.27 &... \\
1059 &23 59 27.56 &+61 00 01.0 &17.21 &1.16 &... & &1263 &00 00 13.05 &+60 56 35.4 &16.28 &1.10 &... \\
1060 &23 59 28.13 &+61 01 08.2 &18.27 &1.23 &... & &1267 &00 00 13.09 &+60 54 50.0 &16.30 &0.79 &... \\
1061 &23 59 27.05 &+60 57 08.2 &16.71 &1.09 &... & &1268 &00 00 13.36 &+60 55 30.2 &16.32 &0.81 &... \\
1062 &23 59 27.87 &+60 58 36.6 &16.25 &0.79 &... & &1270 &00 00 14.81 &+60 58 25.1 &15.14 &0.83 &+0.39 \\
1063 &23 59 27.43 &+60 56 19.0 &16.13 &0.81 &... & &1271 &00 00 12.88 &+60 53 46.9 &16.18 &0.99 &... \\
1064 &23 59 28.87 &+60 59 01.0 &17.09 &1.58 &... & &1272 &00 00 14.21 &+60 56 52.2 &16.68 &0.84 &... \\
1066 &23 59 31.06 &+61 03 11.5 &17.93 &0.73 &... & &1273 &00 00 16.48 &+61 01 21.1 &16.21 &1.21 &... \\
1067 &23 59 29.96 &+61 00 22.6 &16.55 &0.79 &... & &1276 &00 00 15.16 &+60 56 22.6 &16.45 &0.90 &... \\
1068 &23 59 32.09 &+61 03 09.1 &15.75 &0.71 &... & &1277 &00 00 18.04 &+61 02 55.1 &15.96 &1.18 &... \\
1069 &23 59 28.50 &+60 53 37.0 &16.50 &1.90 &... & &1278 &00 00 15.74 &+60 57 35.8 &15.97 &0.73 &... \\
1070 &23 59 30.44 &+60 57 43.2 &16.01 &0.80 &... & &1280 &00 00 15.73 &+60 56 08.6 &14.63 &0.62 &... \\
1071 &23 59 28.90 &+60 53 12.9 &15.76 &0.77 &... & &1281 &00 00 19.00 &+61 03 27.4 &16.65 &1.19 &... \\
1072 &23 59 33.99 &+61 03 29.5 &16.00 &1.24 &... & &1282 &00 00 16.52 &+60 57 18.9 &15.61 &0.66 &... \\
1073 &23 59 34.17 &+61 03 37.8 &16.26 &1.10 &... & &1284 &00 00 17.90 &+61 00 03.0 &17.22 &1.01 &... \\
1074 &23 59 29.68 &+60 52 38.3 &15.41 &0.84 &... & &1285 &00 00 16.72 &+60 56 21.5 &16.29 &0.79 &... \\
1075 &23 59 30.93 &+60 54 47.6 &14.80 &0.79 &... & &1286 &00 00 16.19 &+60 54 46.9 &16.00 &0.80 &... \\
1076 &23 59 32.61 &+60 57 52.0 &17.72 &1.53 &... & &1288 &00 00 18.80 &+61 00 28.6 &16.45 &1.10 &... \\
1077 &23 59 33.73 &+60 59 41.7 &16.10 &1.17 &... & &1289 &00 00 15.18 &+60 52 03.5 &15.89 &0.69 &... \\
1078 &23 59 34.68 &+61 01 48.7 &15.65 &0.98 &... & &1291 &00 00 17.51 &+60 54 33.5 &16.94 &0.93 &... \\
1079 &23 59 31.86 &+60 54 42.7 &17.01 &1.04 &... & &1292 &00 00 19.32 &+60 58 39.6 &17.47 &1.00 &... \\
1080 &23 59 31.86 &+60 54 42.7 &16.97 &1.49 &... & &1296 &00 00 17.29 &+60 53 12.4 &15.17 &0.63 &+0.35 \\
1081 &23 59 32.85 &+60 56 59.5 &16.14 &1.15 &... & &1297 &00 00 18.05 &+60 54 52.8 &18.48 &0.75 &... \\
1082 &23 59 34.12 &+60 59 35.0 &17.33 &1.21 &... & &1300 &00 00 17.98 &+60 53 30.9 &16.95 &0.94 &... \\
1083 &23 59 34.86 &+61 00 37.8 &18.19 &1.32 &... & &1301 &00 00 19.95 &+60 57 43.7 &16.49 &0.83 &... \\
1084 &23 59 35.83 &+61 02 46.5 &17.06 &1.23 &... & &1302 &00 00 22.49 &+61 03 13.7 &17.65 &1.30 &... \\
1085 &23 59 33.94 &+60 58 15.1 &15.44 &1.02 &... & &1303 &00 00 19.95 &+60 56 56.2 &16.21 &0.76 &... \\
1086 &23 59 32.21 &+60 54 02.5 &17.03 &1.15 &... & &1304 &00 00 19.64 &+60 55 42.7 &15.41 &0.82 &+0.40 \\
1087 &23 59 37.51 &+61 03 44.1 &17.09 &1.08 &... & &1305 &00 00 19.28 &+60 54 20.6 &17.33 &0.88 &... \\
1089 &23 59 36.54 &+61 00 02.4 &16.53 &0.98 &... & &1306 &00 00 22.71 &+61 01 56.3 &15.56 &0.97 &... \\
1090 &23 59 36.12 &+60 58 55.6 &17.09 &0.96 &... & &1308 &00 00 20.21 &+60 55 46.7 &16.06 &0.82 &... \\
1091 &23 59 34.93 &+60 55 57.4 &16.78 &0.86 &... & &1309 &00 00 22.89 &+61 01 31.3 &15.37 &0.69 &... \\
1092 &23 59 37.51 &+61 01 02.7 &17.19 &1.11 &... & &1310 &00 00 19.89 &+60 54 14.5 &17.37 &0.94 &... \\
1093 &23 59 34.11 &+60 52 36.4 &17.45 &1.28 &... & &1312 &00 00 23.20 &+61 00 33.3 &16.07 &0.97 &... \\
1094 &23 59 35.83 &+60 56 22.0 &15.86 &0.73 &... & &1313 &00 00 19.55 &+60 52 05.6 &17.12 &1.32 &... \\
1095 &23 59 36.21 &+60 56 32.0 &17.57 &1.22 &... & &1316 &00 00 22.99 &+60 58 18.8 &16.89 &0.91 &... \\
1096 &23 59 37.50 &+60 59 19.8 &16.08 &1.68 &... & &1317 &00 00 22.86 &+60 57 46.9 &16.83 &1.07 &... \\
1097 &23 59 35.15 &+60 52 52.6 &15.92 &1.19 &... & &1318 &00 00 22.56 &+60 56 45.6 &15.96 &0.95 &... \\
1098 &23 59 36.03 &+60 54 59.3 &16.84 &1.25 &... & &1319 &00 00 22.88 &+60 57 13.2 &18.10 &0.97 &... \\
1099 &23 59 39.28 &+61 02 43.6 &16.79 &0.52 &... & &1321 &00 00 25.60 &+61 02 35.0 &17.07 &0.99 &... \\
1100 &23 59 39.64 &+61 02 33.1 &17.46 &0.83 &... & &1322 &00 00 23.71 &+60 57 48.7 &15.95 &1.13 &... \\
1101 &23 59 37.38 &+60 55 12.4 &17.15 &1.26 &... & &1325 &00 00 22.62 &+60 53 54.3 &16.47 &0.84 &... \\
1102 &23 59 38.35 &+60 55 38.7 &16.63 &0.75 &... & &1327 &00 00 23.38 &+60 55 03.2 &16.46 &0.85 &... \\
1103 &23 59 40.10 &+60 59 35.4 &15.86 &1.60 &... & &1329 &00 00 24.44 &+60 55 48.5 &17.27 &1.82 &... \\
1104 &23 59 38.95 &+60 56 32.1 &17.02 &1.16 &... & &1330 &00 00 25.71 &+60 58 11.2 &16.87 &0.95 &... \\
1106 &23 59 41.29 &+61 00 12.6 &15.25 &0.96 &... & &1331 &00 00 27.70 &+61 02 34.9 &16.66 &1.06 &... \\
1108 &23 59 38.51 &+60 52 55.3 &16.57 &0.77 &... & &1334 &00 00 25.39 &+60 56 42.7 &16.73 &1.13 &... \\
1110 &23 59 40.09 &+60 55 57.7 &17.86 &1.22 &... & &1335 &00 00 26.62 &+60 58 55.3 &16.69 &1.00 &... \\
1112 &23 59 43.52 &+61 03 38.3 &16.36 &1.10 &... & &1336 &00 00 28.63 &+61 02 50.2 &17.01 &1.02 &... \\
\hline
\end{tabular}
\end{minipage}
\end{table*}

\setcounter{table}{3}
\begin{table*}
\begin{minipage}{17cm}
\caption{Continued.}
\label{tab4}
\begin{tabular}{ccccccccccccc}
\hline
Star &RA(2000) &DEC(2000) &$V$ &$B-V$ &$U-B$ & &Star &RA(2000) &DEC(2000) &$V$ &$B-V$ &$U-B$ \\
\hline
1113 &23 59 43.33 &+61 02 52.0 &16.41 &0.89 &... & &1337 &00 00 25.72 &+60 56 07.5 &16.60 &0.85 &... \\
1114 &23 59 39.50 &+60 53 42.6 &16.37 &0.97 &... & &1338 &00 00 24.86 &+60 54 02.0 &16.25 &1.18 &... \\
1115 &23 59 40.32 &+60 55 03.8 &15.39 &0.79 &+0.25 & &1340 &00 00 27.13 &+60 58 13.9 &18.06 &1.69 &... \\
1116 &23 59 40.65 &+60 54 36.6 &16.49 &0.82 &... & &1341 &00 00 26.64 &+60 56 47.9 &16.46 &1.01 &... \\
1117 &23 59 41.73 &+60 56 35.4 &15.32 &0.86 &... & &1343 &00 00 26.18 &+60 54 16.6 &16.25 &1.04 &... \\
1119 &23 59 43.50 &+60 58 13.5 &17.14 &1.22 &... & &1344 &00 00 29.86 &+61 01 29.8 &16.83 &1.34 &... \\
1120 &23 59 45.69 &+61 02 26.1 &15.32 &1.07 &... & &1345 &00 00 25.91 &+60 52 38.3 &17.19 &0.86 &... \\
1121 &23 59 42.53 &+60 53 55.6 &16.28 &0.72 &... & &1346 &00 00 29.97 &+61 01 08.2 &17.60 &0.84 &... \\
1122 &23 59 46.20 &+61 02 01.7 &17.52 &1.25 &... & &1347 &00 00 27.25 &+60 54 15.9 &16.34 &0.97 &... \\
1123 &23 59 43.41 &+60 55 23.6 &17.04 &0.84 &... & &1348 &00 00 28.62 &+60 56 47.1 &16.96 &1.08 &... \\
1125 &23 59 44.33 &+60 56 58.6 &16.72 &0.89 &... & &1349 &00 00 28.30 &+60 55 49.0 &15.64 &0.72 &... \\
1126 &23 59 43.56 &+60 55 04.0 &16.21 &2.03 &... & &1350 &00 00 27.82 &+60 53 48.3 &16.70 &1.70 &... \\
1127 &23 59 47.02 &+61 03 03.1 &17.31 &0.91 &... & &1351 &00 00 27.71 &+60 53 14.2 &15.58 &0.80 &+0.36 \\
1128 &23 59 46.62 &+61 01 32.7 &15.44 &0.84 &... & &1352 &00 00 31.85 &+61 02 28.5 &17.52 &1.11 &... \\
1129 &23 59 46.50 &+60 59 05.0 &16.56 &1.29 &... & &1353 &00 00 31.05 &+61 00 30.3 &16.71 &0.95 &... \\
1131 &23 59 45.21 &+60 55 29.1 &16.97 &0.88 &... & &1354 &00 00 32.37 &+61 03 24.7 &17.55 &1.31 &... \\
1132 &23 59 48.63 &+61 02 17.7 &16.29 &1.05 &... & &1355 &00 00 28.37 &+60 53 26.7 &18.03 &1.30 &... \\
1133 &23 59 45.25 &+60 53 56.4 &17.55 &1.09 &... & &1356 &00 00 31.96 &+61 00 51.7 &16.23 &0.84 &... \\
1134 &23 59 45.19 &+60 53 10.1 &16.73 &1.04 &... & &1357 &00 00 29.40 &+60 54 53.3 &17.67 &1.61 &... \\
1135 &23 59 46.91 &+60 57 15.7 &16.02 &0.86 &... & &1358 &00 00 33.10 &+61 02 31.4 &15.83 &1.57 &... \\
1136 &23 59 49.94 &+61 02 53.1 &17.51 &1.62 &... & &1360 &00 00 31.91 &+60 55 55.0 &16.47 &1.13 &... \\
1137 &23 59 46.12 &+60 53 22.2 &16.08 &0.95 &... & &1361 &00 00 35.02 &+61 02 12.6 &17.79 &0.75 &... \\
1138 &23 59 49.74 &+61 00 38.8 &15.68 &0.70 &... & &1362 &00 00 34.55 &+61 00 38.4 &17.98 &1.19 &... \\
1140 &23 59 50.67 &+61 01 41.2 &15.58 &1.08 &... & &1366 &00 00 31.86 &+60 51 48.7 &15.76 &0.89 &... \\
1141 &23 59 48.83 &+60 56 27.6 &15.59 &0.43 &... & &1368 &00 00 35.45 &+60 59 16.8 &17.76 &1.30 &... \\
1142 &23 59 48.58 &+60 54 55.9 &16.75 &0.75 &... & &1369 &00 00 34.42 &+60 56 17.0 &16.72 &1.68 &... \\
1143 &23 59 48.15 &+60 52 45.3 &17.97 &1.15 &... & &1370 &00 00 34.37 &+60 55 24.0 &16.60 &1.62 &... \\
1145 &23 59 52.16 &+61 01 56.9 &15.26 &0.67 &... & &1371 &00 00 34.24 &+60 53 47.0 &16.69 &1.01 &... \\
1146 &23 59 49.31 &+60 54 51.9 &16.66 &1.48 &... & &1372 &00 00 34.82 &+60 54 43.0 &16.20 &0.71 &... \\
1147 &23 59 49.43 &+60 54 30.0 &17.56 &1.22 &... & &1373 &00 00 34.78 &+60 53 52.4 &15.45 &1.00 &... \\
1148 &23 59 51.90 &+60 59 21.9 &16.51 &1.08 &... & &1374 &00 00 36.03 &+60 54 42.1 &17.24 &1.15 &... \\
1149 &23 59 51.37 &+60 58 06.1 &15.19 &0.82 &... & &1375 &00 00 39.55 &+61 02 38.1 &16.98 &1.15 &... \\
1152 &23 59 53.67 &+61 02 36.0 &17.47 &1.18 &... & &1376 &00 00 37.12 &+60 57 02.4 &16.00 &1.08 &... \\
1153 &23 59 54.32 &+61 03 11.1 &16.90 &0.82 &... & &1377 &00 00 39.04 &+61 00 55.9 &16.67 &1.34 &... \\
1154 &23 59 51.16 &+60 55 14.9 &17.80 &1.65 &... & &1379 &00 00 37.67 &+60 57 59.9 &15.55 &1.02 &... \\
1155 &23 59 52.79 &+60 58 19.0 &17.38 &0.83 &... & &1381 &00 00 40.36 &+61 02 47.1 &15.80 &1.67 &... \\
1157 &23 59 53.48 &+60 59 19.6 &15.41 &0.78 &... & &1382 &00 00 40.89 &+61 03 10.5 &15.40 &1.48 &... \\
1158 &23 59 53.70 &+60 58 05.2 &15.70 &1.13 &... & &1383 &00 00 37.34 &+60 55 18.2 &17.14 &0.99 &... \\
1159 &23 59 53.37 &+60 56 31.1 &16.86 &0.86 &... & &1384 &00 00 38.06 &+60 56 22.3 &17.04 &1.35 &... \\
1161 &23 59 53.61 &+60 56 42.3 &16.31 &0.64 &... & &1385 &00 00 39.47 &+60 59 25.0 &17.78 &1.05 &... \\
1162 &23 59 53.84 &+60 56 47.8 &17.30 &1.07 &... & &1386 &00 00 36.10 &+60 51 55.4 &17.95 &0.91 &... \\
1163 &23 59 55.67 &+61 01 05.6 &16.52 &2.00 &... & &1387 &00 00 41.30 &+61 02 53.3 &15.70 &1.15 &... \\
1164 &23 59 59.66 &+60 58 23.2 &15.94 &0.88 &... & &1388 &00 00 36.89 &+60 52 34.9 &16.71 &0.75 &... \\
1165 &23 59 55.58 &+60 58 07.6 &15.57 &0.97 &... & &1390 &00 00 39.24 &+60 56 41.0 &16.81 &1.00 &... \\
1166 &23 59 55.72 &+60 58 02.6 &16.54 &1.06 &... & &1391 &00 00 39.73 &+60 56 18.0 &17.05 &0.98 &... \\
1167 &23 59 56.69 &+61 00 12.9 &17.40 &1.22 &... & &1392 &00 00 40.08 &+60 56 38.0 &15.57 &1.16 &... \\
1170 &23 59 55.08 &+60 52 26.3 &16.83 &1.07 &... & &1393 &00 00 39.00 &+60 54 19.5 &16.79 &0.84 &... \\
1171 &23 59 56.83 &+60 55 47.9 &16.50 &0.80 &... & &1395 &00 00 39.72 &+60 54 38.8 &16.34 &1.74 &... \\
1172 &23 59 59.55 &+61 01 53.2 &16.69 &0.94 &... & &1396 &00 00 43.80 &+61 02 59.9 &15.14 &1.08 &... \\
1173 &23 59 57.85 &+60 57 25.1 &16.45 &0.98 &... & &1397 &00 00 41.17 &+60 54 50.4 &15.25 &1.12 &... \\
1174 &00 00 00.65 &+61 03 36.3 &15.55 &1.06 &... & &1398 &00 00 40.57 &+60 52 41.2 &17.19 &0.86 &... \\
1176 &23 59 56.51 &+60 52 55.0 &16.74 &1.11 &... & &1399 &00 00 38.87 &+60 53 10.0 &16.01 &0.68 &... \\
1177 &00 00 00.62 &+61 01 12.0 &17.52 &1.34 &... & &1400 &00 00 43.57 &+60 58 21.3 &15.45 &0.76 &... \\
1178 &00 00 00.20 &+60 59 46.3 &17.25 &0.81 &... & &1401 &00 00 44.40 &+60 59 36.9 &18.65 &1.07 &... \\
1179 &00 00 01.88 &+61 03 03.4 &15.38 &1.07 &... & &1402 &00 00 43.42 &+60 56 29.0 &15.88 &0.82 &... \\
1182 &00 00 00.15 &+60 57 08.5 &15.28 &1.01 &... & &1403 &00 00 42.83 &+60 54 31.8 &15.47 &0.79 &... \\
1183 &00 00 00.26 &+60 56 27.9 &16.88 &1.92 &... & &1404 &00 00 43.86 &+60 54 44.7 &16.30 &0.96 &... \\
1184 &23 59 58.63 &+60 52 34.4 &16.41 &1.09 &... & &1405 &00 00 43.28 &+60 52 49.2 &17.41 &0.98 &... \\
1185 &00 00 00.30 &+60 56 02.3 &16.11 &0.75 &... & &1406 &00 00 44.94 &+60 55 57.8 &16.27 &1.06 &... \\
1188 &23 59 58.82 &+60 52 06.4 &15.79 &0.73 &... & &1407 &00 00 44.69 &+60 55 09.5 &16.07 &1.10 &... \\
1189 &23 59 59.50 &+60 53 34.1 &17.02 &1.53 &... & &1409 &00 00 44.26 &+60 52 52.3 &15.79 &0.76 &... \\
1190 &00 00 03.53 &+61 02 52.6 &16.24 &1.22 &... & &1410 &00 00 44.22 &+60 52 20.0 &16.66 &1.32 &... \\
1191 &00 00 00.29 &+60 54 38.8 &15.95 &0.94 &... & &1411 &00 00 49.52 &+61 02 30.2 &15.75 &1.66 &... \\
\hline
\end{tabular}
\end{minipage}
\end{table*}

\end{document}